\documentclass[fleqn,usenatbib]{mnras}

\usepackage{newtxtext,newtxmath}
\usepackage{ae,aecompl}
\usepackage{{booktabs}}
\usepackage[T1]{fontenc}

\DeclareRobustCommand{\VAN}[3]{#2}
\let\VANthebibliography\thebibliography
\def\thebibliography{\DeclareRobustCommand{\VAN}[3]{##3}\VANthebibliography}

\usepackage{graphicx}	% Including figure files
\usepackage{amsmath}	% Advanced maths commands
\usepackage{bbm}
\usepackage{graphicx}
\usepackage{lipsum}
\usepackage{multirow}
\usepackage{natbib}
\usepackage{subcaption}
\usepackage{soul}
\usepackage[separate-uncertainty=true]{siunitx}
\usepackage[dvipsnames]{xcolor}   
\usepackage{bbold}
\usepackage[normalem]{ulem}
\usepackage{orcidlink}
\newcommand{\Cell}{\ensuremath{\mathcal{C}_\ell}}

\newcommand{\lcdm}{$\Lambda$CDM }

\definecolor{deepmagenta}{rgb}{0.8, 0.0, 0.8}
\definecolor{ballblue}{rgb}{0.13, 0.67, 0.8}

\definecolor{RedWine}{rgb}{0.743,0,0}

\definecolor{verde}{rgb}{0,0.5,0}

\author[K. Lodha et al. ]{Kushal Lodha$^{1,2}$\orcidlink{0009-0004-2558-5655}, Lucas Pinol$^{3}$\orcidlink{0000-0002-2218-5929}, Savvas Nesseris$^{3}$\orcidlink{0000-0002-0567-0324}, Arman Shafieloo$^{1,2}$\thanks{E-mail: \href{mailto:shafieloo@kasi.re.kr }{shafieloo@kasi.re.kr}}\orcidlink{0000-0001-6815-0337}, Wuhyun Sohn$^{1}$\orcidlink{0000-0002-6039-8247},\newauthor and Matteo Fasiello$^{3}$\orcidlink{0000-0002-2532-5202}  \\ %({\verde please put me last given the limited contribution})\, \\
$^{1}$  Korea Astronomy and Space Science Institute, Daejeon 34055, Republic of Korea\\
$^{2}$ KASI Campus, University of Science and Technology, Daejeon 34113, Republic of Korea\\
$^{3}$ Instituto de F\'isica Te\'orica IFT UAM-CSIC, Universidad Auton\'oma de Madrid, Cantoblanco, E-28049 Madrid, Spain}

\title[Local features in primordial spectrum]{Searching for local features in primordial power spectrum using genetic algorithms}

\date{Accepted 2024 March 12. Received 2024 March 4; in original form 2023 August 24}
\pubyear{2024}
\volume{530}
\pagerange{1424–1435}
\journal{{MNRAS} {\bf 530}, 1-12
  (2024) \hfill \href{https://doi.org/10.1093/mnras/stae803}{https://doi.org/10.1093/mnras/stae803}}
\begin{document}
\label{firstpage}
\pagerange{\pageref{firstpage}--\pageref{lastpage}}
\maketitle

\begin{abstract}
We present a novel methodology for exploring local features directly in the primordial power spectrum using a genetic algorithm pipeline coupled with a Boltzmann solver and Cosmic Microwave Background data (CMB).
After testing the robustness of our pipeline using mock data, we apply it to the latest CMB data, including Planck 2018 and CamSpec PR4.
Our model-independent approach provides an analytical reconstruction of the power spectra that best fits the data, with the unsupervised machine learning algorithm exploring a functional space built off simple `grammar' functions.
We find significant improvements upon the simple power-law behaviour, by $\Delta \chi^2 \lesssim -21$, consistently with more traditional model-based approaches.
These best-fits always address both the low-$\ell$ anomaly in the TT spectrum and the residual high-$\ell$ oscillations in the TT, TE and EE spectra.
The proposed pipeline provides an adaptable tool for exploring features in the primordial power spectrum in a model-independent way,
providing valuable hints to theorists for constructing viable inflationary models that are consistent with the current and upcoming CMB surveys.

\end{abstract}


%%%%%%%%%%%%%%%%%%%%%%%%%%%%%%%%%%%%%%%%%%%%%%%%%%

%%%%%%%%%%%%%%%%% BODY OF PAPER %%%%%%%%%%%%%%%%%%

\section{Introduction}

The most cogent case for inflation comes from its ability to make sense of the homogeneity at very large angular scales in the Cosmic Microwave Background. These otherwise puzzling measurements are readily explained by an accelerated expansion that took place very early in the Universe's evolution. Additionally, cosmic inflation provides a natural mechanism for quantum fluctuations, which give rise to density fluctuations that eventually grow into the large-scale structure we see today.

The simplest inflationary paradigm delivering a viable cosmology consists of a scalar field slowly rolling down an almost flat potential, resulting in adiabatic scalar perturbations with a nearly scale-invariant power spectrum.
The conventional way to characterise this scalar power spectrum is via a power law of the form:
\begin{equation}
    P(k) = A_{\mathrm s} \left(\frac{k}{k_*}\right)^{n_{\mathrm s} - 1 },
\end{equation}
where $A_{\mathrm s}$, $n_{\mathrm s}$ and  $k_* = 0.05\,\mathrm{Mpc}^{-1}$ are respectively the amplitude, spectral index, and pivot scale of the primordial spectrum.
Similarly, one may show that, in the simplest scenario, tensor fluctuations are produced from vacuum fluctuations and engender a primordial tensor power spectrum that is also well described by a power law  $P_t(k) = A_\mathrm{t} \left(k/k_*\right)^{n_{\mathrm t}}$.
Planck's 2018 data analysis has accurately measured the scalar power law spectrum to be slightly red-tilted and placed constraints on the maximum amount of tensor modes at CMB scales. Assuming the power law, such measurements reveal that the inflationary models must correctly predict $\ln(10^{10} \ A_{\mathrm s}) = 3.044 \pm 0.014$, $n_{\mathrm s}=0.9649 \pm 0.0042$ at 68\% confidence level, and $r_{0.002} \simeq A_{\mathrm t}/A_{\mathrm s} < 0.10 $ at $95\%$ confidence level \citep{PlanckCollaboration2018power}, with even tighter constraints on primordial tensor modes when combined with CMB data from the BICEP-Keck collaboration: $r_{0.05}<0.035$~\citep{BICEPKeck:2022mhb}.

{The motivation for exploring features beyond this simple power law in the primordial scalar power spectrum is manifold. Firstly, they provide valuable insights into the dynamics of inflation itself.
Indeed, several compelling models of inflation comprise transient deviations from the slow-roll dynamics. These features in the inflationary trajectory may be due to localised bumps, dips, saddles points, etc., in the scalar potential or its derivatives or to a sudden turn in multi-dimensional field space, resulting in some features and oscillations in the scalar power spectrum (see, e.g. \cite{Starobinsky:1992ts,Chung:1999ve,Adams2001InflationaryStep,Gong_2005,Joy_2009,Ach_carro_2011,Achucarro:2014msa,Mizuno:2014jja,Braglia:2020fms,Fumagalli:2020nvq,Dalianis:2021iig}).
This family of features is called sharp features, for the deviation from smoother slow-roll dynamics is only transient. Another family of features, so-called resonant features, may be produced by tiny oscillations overimprinted on the slow-roll potential, or by the oscillations of an excited massive scalar field in a multifield context \citep{Chen:2011zf,Battefeld_2013, Gao_2013, Chen_2014, Meerburg_2014,Palma_2015}.
Contrary to sharp features, resonant features result in oscillations in $\log k$, which may be local or global.
Other kinds of features may be encountered, leading to peaks and troughs in the scalar power spectrum, broken power laws or plateaus, see~\citep{Slosar_2019} for a recent review on primordial features. Thus, identifying and characterising features in the primordial power spectrum is then a key step towards gaining a deeper understanding of the inflationary mechanism, its microphysics and particle content.

Our analysis is further motivated by the fact that, despite the success of the simple power law power spectrum, there are several issues with the standard \lcdm model, which include the $A_{\mathrm L}$ anomaly \citep{PlanckCollaboration2018parameters}, the Hubble \citep{Riess_2019} and $S_8$  tensions (see \cite{Valentino_2021_fs8} for reviews).\\ 
The phenomenological lensing consistency parameter,  $A_\mathrm{L}$, was introduced to test the consistency of the amount of lensing-induced smoothing in the angular power spectrum data with the theoretical prediction of $A_\mathrm{L}=1$ \citep{Calabrese_2008_WL}. Analyses of Planck data have a preference for $A_\mathrm{L}>1$ at the 3$\sigma$ level when considering only power spectrum data, which decreases to $2\sigma$ when including lensing reconstructions \citep{PlanckCollaboration2018inflation}. The presence of $A_\mathrm{L} > 1$ ``anomaly'' suggests additional smoothing of spectra compared to the predictions of the standard cosmological model. The ``Hubble tension'' is a 3.4$\sigma$ inconsistency between the measurement of the current expansion rate of the Universe in our close-by environment and the one extrapolated from precision CMB observations at the time of photon decoupling. The presence of features in the primordial power spectrum could potentially offer insights into the underlying causes of these tensions by inducing shifts in other cosmological parameters and reconciling the existing tensions within the standard \lcdm paradigm (see , e.g.\cite{Hazra_2019, Hazra_2022, Ballardini_2022, Antony_2023}).

We study the form of the primordial power spectrum using the CMB temperature and polarisation data from Planck which offers the most sensitive avenue to search for these signatures. At the CMB scales, the modes exhibit sufficient linearity, allowing us to utilise forward modelling to isolate the primordial information. However, the original signal is significantly suppressed by Silk damping resulting from photon diffusion, particularly beyond $k \gtrsim \mathcal{O}(10^{-1})$ Mpc$^{-1}$, and tends to be overshadowed by foregrounds \citep{Chluba_2015}. Besides the CMB, large-scale structure (LSS) surveys offer another opportunity to probe deviations from a featureless primordial power spectrum. Despite the recent advancements in handling non-linearities for global (not localised in k) oscillatory features \citep{Beutler_2019,mergulhao2023primordial}, the modelling uncertainties in the nonlinear regime constrain our ability to comprehensively test arbitrary forms of features using large-scale structure data \citep{Hu:2014aua, Hunt2015SearchDatasets, Benetti_2016, Zeng_2019}.

In order to search for features in the primordial spectrum, two fundamental approaches may be considered: parametric and non-parametric ones.

Parametric approaches involve extending the power law primordial spectrum with a specific functional template and fitting the parameters of this extended model to observational data. This includes a model with varying amplitude and tilt of density perturbations as a power law of the scale of the perturbation \citep{Cline_2006,PlanckCollaboration2018inflation}. Another subclass are features where deviations from a smooth power spectrum are confined to a particular range of scales. These ``local features'' include broken power-law spectrum, sharp peaks and dips, and resonant oscillations in the spectrum at a specific scale \citep{Chluba_2015,Slosar_2019}.  Parametric approaches have the advantage of being interpretable and computationally efficient to test with the data. However, the downside is that they may not be flexible enough to capture more complex features in the primordial spectrum.

Non-parametric approaches, on the other hand, do not assume a specific functional form for the primordial spectrum. Previous works have studied alternative forms to the primordial power spectrum by varying its amplitudes in different wavenumber bins \citep{Wang1999CosmologyModels,Hannestad2001ReconstructingData,Hannestad2004ReconstructingAlgorithm,Elgaroy2002} or at some nodes (knots) \citep{Bridle2003ReconstructingSpectrum,Bridges2009BayesianSpectrum,Sealfon_2005,Verde2008spline,Peiris2010spline,Vazquez2012ModelSpectrum,Aslanyan2014TheSpectrum,Abazajian2014TheSpectrum,PlanckCollaboration2015inflation,PlanckCollaboration2018inflation,Handley2019BayesianData}, searched for localised feature using wavelets \citep{Mukherjee2003DirectSpectrum,Mukherjee2005PrimordialReconstruction,Paykari:2012}, and obtained free-form reconstructions via maximising the likelihood with some regularisation terms \citep{Tegmark2002SeparatingBox,Tocchini-Valentini2005ASpectrum,Tocchini-Valentini2006Non-parametricData,Nagata2009Band-powerMethod,Hunt2014ReconstructionSets,Hunt2015SearchDatasets,Gauthier2012ReconstructingCMB,PlanckCollaboration2013inflation,PlanckCollaboration2015inflation,PlanckCollaboration2018inflation}, the cosmic inversion method \citep{Matsumiya2002CosmicAnisotropy,Matsumiya2003CosmicData,Kogo2004ReconstructingMethod}, principal component analysis \citep{Hu2004principal,Leach_2006}, and deconvolution algorithms \citep{Shafieloo2004MRL,Shafieloo2007FeaturesAnalysis,Shafieloo2008EstimationData,Nicholson2009ReconstructionExperiments,Gibelyou2010DetectabilityDistribution,Hamann2010FeaturesAnalysis,Hazra2013wmapMRL,Hazra2013CosmologicalSpectrum,Hazra2014PlanckMRL,Chandra2021PrimordialSpectrum,Sohn_2922}. Non-parametric methods have the advantage of being more flexible and able to capture a wider range of features in the data with little theoretical bias; the downside is that they can be computationally expensive and potentially difficult to interpret their statistical significance} because they often involve a high level of functional freedom.

Genetic algorithms (GAs) are meta-heuristic optimisation techniques that are inspired by the process of natural evolution. Since the introduction of GA in the mid-1970s \citep{Holland_1975}, they have been widely adopted in various fields, such as engineering, economics, accelerator physics and cosmology, to search for the optimal solution to a particular problem. 
GAs have been coupled with Wilson-Devinney (W-D) code to optimise the light curve of eclipsing binaries \citep{Metcalfe_1999} and with radiative transfer codes to fit dust spectra from AGB stars \citep{Baiser_2010}.
In the context of cosmology, GAs were used for reconstructing the expansion history of the Universe in a model-independent fashion \citep{Bogdanos_2009,Nesseris:2012tt,Nesseris_2013,gangopadhyay2023phantom}, to test the validity of \lcdm in combination with Om-statistics \citep{Nesseris:2010ep}, growth rates \citep{Arjona_2022}, constraining deviations from general relativity \citep{Alestas_2022} and finding approximate analytical functions for transfer functions \citep{Orjuela-Quintana22} and sound horizon at drag epoch \citep{Aizpuru_2021}.

In this work, we present a novel approach to directly study features in the primordial power spectrum using a genetic algorithm. Our method is designed to explore this high-dimensional feature space by introducing combinations of local and global features in the primordial power spectrum, utilising a grammar that captures these essential aspects.
GAs have been recently applied to the search for primordial features in the scalar potential of a single scalar field driving inflation ~\citep{Kamerkar_2022, Abel_2022}, providing a complementary approach to the one we shall pursue here. To validate our approach, we test it on mock data before applying it to the Planck 2018 likelihood and the latest release from CamSpec PR4. Our results demonstrate the effectiveness of our proposed methodology in identifying essential features in the primordial power spectrum, which can provide valuable insights into our understanding of the early Universe.

The paper is organised as follows: In Section 2, we provide a concise overview of the genetic algorithm and pipeline utilised to search for features that we validate with mocks. Section 3 describes the datasets used for the analysis and presents the results. We finally conclude the paper with a summary and discussion of possible future directions in Section 4.

\section{Methodology}

To compute the angular power spectrum of the CMB, one needs to solve the Einstein-Boltzmann equations, which describe the evolution of photon-baryon perturbations in the early Universe.
This hierarchy of ordinary differential equations can be solved numerically using cosmological Boltzmann solvers, which take the primordial power spectrum as an input.
The solution of the Boltzmann equation yields transfer functions that describe the evolution of the baryon-photon plasma and the growth of the perturbations. The transfer functions depend on various cosmological parameters, such as the density of dark matter, baryons, and dark energy, as well as the Hubble constant and the amplitude of primordial perturbations. We can calculate the CMB temperature and polarisation anisotropies in terms of the primordial power spectrum using the relation:
\begin{equation}
C^{XY}_{\ell} \propto 4\pi\int \frac{dk}{k}\,P(k)\,\Big[T_{\ell}^{X}(k)\,T_{\ell}^{Y}(k)\Big],    
\end{equation}
where $P(k)$ is the primordial scalar power spectrum, $T^X_{\ell}(k)$ is the transfer function where $X=T, E$ correspond to the temperature and E-mode polarisation, respectively, and $\ell$ is the multipole moment. While the constraints on primordial $B$ modes from experiments such as BICEP-Keck \citep{BICEPKeck:2022mhb} can provide valuable information on the tensor counterpart of the primordial scalar power spectrum, the fact that Planck 18 data used in this work does not include $B$ modes and only provides a weak constraint on primordial tensors, $r_{0.002}<0.10$, motivates us to leave reconstruction of tensor modes for future works.

\subsection{Genetic Algorithms}
 
In our approach, we represent the desired power spectrum as chromosomes, akin to the composite structure found in genetics, while individual features can be seen as analogous to `genes.' A fitness function analogous to a loss function in machine learning acts as a mapping between chromosomes and the properties of the desired solution. It serves as a means to assess the proximity of a candidate solution to the available data. In our case, the fitness function is defined as the sum of $\chi^2$ values derived from different likelihoods. Through the utilisation of genetic algorithms, we aim to optimise the selection of features and iteratively refine the solution space to identify the most promising configurations that best agree with the given data.

The basic methodology of a GA can be summarised as follows:
\begin{itemize}
    \item Initialisation: A population of individuals is randomly generated, each representing a potential solution to the optimisation problem.
    \item Fitness Evaluation: The fitness of each individual in the population is evaluated using a fitness function, which measures how well the individual solves the optimisation problem.
    \item Selection: The fittest individuals are selected from the population to produce the next generation of individuals. The likelihood of reproduction is based on the fitness scores of the individuals.
    \item Crossover: The selected individuals are combined to form a new generation of offspring. This is done by taking bits of genetic information from each parent and combining them to form a new individual.
    \item Mutation: A small percentage of the offspring are randomly mutated to introduce new genetic diversity into the population.
    \item Termination: The algorithm stops when a specified number of generations is reached, or the individual fulfils pre-defined fitness criteria.  
\end{itemize}

\begin{table}
\centering
\caption{The priors for the Eq.~\eqref{eq:feature_gen} that determine the amplitude, shape, width and position of the feature. The initial population for GA analysis is created by randomly sampling this parameter space. These ranges are explicitly enforced by means of constrained crossover and mutation operators. }
\begin{tabular}{|c|c|}
\hline
Parameter & Prior Range                                                                                   \\
\hline
A       & [-2,4]                                                                                    \\
B       & [0, len(grammar)]  \\
C       & [$\log(0.8)$, $\log(10)$]                                                                                              \\
D       & [$\log(0.01)$, $\log(5)$] \\
\hline
\end{tabular}
\label{Tab:prior}
\end{table}

\subsection{Pipeline}

We first generate four sets of four random numbers (A, B, C, D), each from a pre-defined prior range as presented in Table~\ref{Tab:prior}. Each set represents a distinct feature, comprising the amplitude, shape from grammar, scale, and position of the feature. These features are linearly combined to generate an individual using Eq~\eqref{eq:feature_gen}. 
Utilizing the Sympy library \citep{sympy_2017}, we transform each individual into an analytical expression, which is then incorporated into the primordial module of CLASS \citep{Lesgourgues:2011re} to calculate the corresponding angular power spectrum ($C_{\ell}$s). Subsequently, the computed $C_{\ell}$s are passed to `clik' \citep{PlanckCollaboration2018power} to compute the log-likelihood value. Due to the computational complexity of our approach, we keep the background cosmological and nuisance parameters fixed to the BOBQYA \citep{Powell2009TheBA} best-fit values from the vanilla \lcdm model.

The deviation to power law spectra is modelled  by 
\begin{equation}
\label{eq: def GA PS}
\frac{P_{\mathrm{GA}}}{P_{0}}(x) = |1 + F(x)|\,,
\end{equation}
where $x = \log\left(\frac{k}{k_*}\right)$ and $F(x)$ is the linear combination of features in logarithmic space, in our case, we limit its length to four terms, see Eq.~\eqref{eq:feature_gen}. The selection of combining four features linearly in our methodology is based on preliminary trial runs. If we include fewer features, the GA's effectiveness would be diminished, see \citep{Kamerkar_2022}. Conversely, incorporating an excessive number of features could hinder optimisation. These trials aimed to strike a suitable compromise, providing sufficient flexibility in the P(k) space while minimising the risk of poor convergence. 
The presence of the modulus in Eq.~\eqref{eq: def GA PS} is motivated by our will always to have P(k) positive and avoid unphysical solutions with negative power. Note that the GA looks for features that are functions of the $\log(k)$ instead of $k$ to better parametrise the relative power changes across many scales.

So, the overall function reconstructed by the GA is given by
\begin{equation}
\label{eq:feature_gen}
    F(x) =  \sum_{i=1}^{4} f_i(x)\,,
\end{equation}

where
\begin{equation}
    f(x)= 
\begin{cases}
    A\sqrt{C}G_B(C(x-D)) \ e^{-(C(x-D))^2},& \text{\tiny if Local Grammar}\\
    A \ G_B(C(x-D)),              & \text{\tiny Global Grammar}
    %\label{eq:feature_gen}
\end{cases}
\end{equation}
and $G_B(x)$ are functions in the grammar as described in Table~\ref{tab:grammar}.

These two classes, local and global, are motivated by inflationary scenarios where the deviation from the slow-roll dynamics is nearly instantaneous, or on the contrary, present during the whole evolution.
For example, if a slow-roll potential is superimposed by a tiny but persistent, periodic modulation, $V_\mathrm{slow-roll}(\phi) \rightarrow V_\mathrm{slow-roll}(\phi) [1+ \alpha \mathrm{sin}(\phi/\Lambda)]$ with $\alpha \ll 1$ and $\Lambda \ll M_\mathrm{Pl}$, the power-law primordial spectrum is modified to feature a global oscillatory component, with oscillations linear in $\log k$ space across all scales~\citep{Chen:2008wn}, therefore also justifying the use of the variable $x$.

\begin{table}
\caption{Overview of the grammar used in GA analysis, including global and local classes, along with the corresponding functions associated with each class. The global class encompasses trigonometric functions and polynomials, while the local class comprises various combinations of trigonometric, polynomial, and higher-order trigonometric functions. }
\begin{tabular}{|c|c|c|}
\hline
Grammar Type & Class                      & Functions                                                                                       \\
\hline
Global       & Trigonometric              & $\sin{x}$,$\cos{x}$                                                                                   \\
             & Polynomials                & $x$                                                                                               \\
\hline
Local        & Trigonometric              & $\sin{x}$, $\cos{x}$                                                                                  \\
             & Trigonometric/Polynomial   &  $\sin{x}/x$ , $\cos{x}/x$                                                                             \\
             &                            & $x  \sin{x}$ , x $\cdot$ cos(x)                                                                              \\
             & Higher Order Trigonometric & \begin{tabular}[c]{@{}l@{}}sin(2x), cos(2x),\\ sin(4x), cos(4x)\\ sin(8x), cos(8x)  \end{tabular}\\
             & Special &  $\mathbb{1}$, $\mathbb{0}$ \\ 

\hline
\end{tabular}
\label{tab:grammar}
\end{table}

Another possibility is that the oscillatory component is localised around a particular region of the inflationary potential or that the resonant feature comes from the temporary excitation of a heavier degree of freedom, such as background oscillations of a second scalar field in multifield inflation, see, e.g.,~\citep{Chen:2014cwa} for explicit two-field realisations.
Such a localised, so-called resonant feature results in oscillations in $\log k$ space again, but this time with a finite extension, explaining our local grammar.
We comment that yet another possibility is one of the so-called sharp features, where the transient deviation from a slow roll is very short but very violent, and where no particular oscillation is present in the background, see~\citep{Slosar_2019, martin_2023_encyclopaedia} for a recent review on primordial features and their relations to inflationary physics.
Sharp features result in localised oscillations linear in $k$ space and are therefore not encompassed by our present study; we leave for future work the search for sharp features in the CMB data with GA. The fitness of each individual is then evaluated based on its negative log-likelihood ($-\log \mathcal{L}$) value. To form a mating pool, we employ a binary tournament selection process where the fittest individual receives the maximum number of copies. From this pool, we randomly select two individuals for each pairing, generating a total of $N_{\mathrm{pop}} \times p_{\mathrm{c}}$ offsprings via a crossover, with the crossover individuals being mutated with a probability of $p_{\mathrm{m}}$. We set crossover and mutation probability to 0.8 and 0.4, respectively. The remaining individuals ($N_{\mathrm{pop}} \times (1- p_{\mathrm{c}}$)) for the next generation are selected from the fittest individuals of the previous generations, ensuring their continued contribution to the evolving population. 

For the crossover operator, we adopt a combination of one-point crossover and a bounded version of Simulated Binary crossover (SBC) crossover \citep{Deb_1995simulated} for our case through which values of $A_i, C_i$ and $D_i$ are generated by
\begin{align}
\begin{split}    
X'_j &= \frac{1}{2} \left[ (1+ \beta(u))X_j + (1-\beta(u)) X_{j+1} \right], \\
X'_{j+1} &= \frac{1}{2} \left[ (1 - \beta(u))X_j + (1+\beta(u)) X_{j+1} \right]
\end{split}
\end{align}
where $\beta(u)$ and u is a random number in the range $\in [0,1]$ 
\begin{equation}
    \beta(u)= 
\begin{cases}
    \left(2*(1-u)\right)^{-1.0/(\eta_c+1)},& \text{if} \ u>0.5 \\
    \left(2*(u)\right)^{-1.0/(\eta_c+1)},              & u \leq 0.5
\end{cases}
\end{equation}

If a parameter exceeds the upper or lower bounds of the prior, it is clipped to be within the permissible range. The $\eta_c$ is a hyper-parameter which can be tuned to improve the performance, but we fix $\eta_{\mathrm{c}}$ to 2. The motivation for using SBC comes from its ability to consistently generate unique offspring from the same set of parents.  

For mutation, we employ the Non Uniform Mutation (NUM) operator \citep{Michalewicz_1996GeneticA}.
Specifically, we randomly select a feature, say $f_i(x)$, from the individual and tweak one of its parameters. We allow the mutation operator to alter the amplitude ($A_i$), width ($C_i$) or position ($D_i$) of the feature but leave the shape unaltered.
Specifically, we draw two random numbers $(u,v)$ in the range $\in [0,1]$. The first random number ($v$) determines the sign, while the second one ($u$) determines the magnitude. 

\begin{equation}
    X_i'= 
\begin{cases}
   X_i + \text{range(X)}(1-u^{\eta_m}),& \text{if} \ v > 0.5 \\
  X_i - \text{range(X)}(1-u^{\eta_m}), &  v \leq 0.5
\end{cases}
\end{equation}
Here, $X_i$ represents the parameter's original value in consideration, while $X_i'$ denotes the mutated value. The term $\text{range}(X)$ refers to the parameter's allowed range of prior values. The mutation index, denoted as $\eta_m$, is fixed at 2 for all further analyses.

We evolved our individuals for 800 generations since there was no significant improvement in the $\chi^2$ during our initial runs. 
We reiterate the fact that GAs are stochastic in nature and do not guarantee finding the global optimum. The quality of the solution is contingent upon factors such as grammar, hyper-parameters, as well as the choice of crossover and mutation operators utilised.

\subsection{Validation with Mocks}

We borrow the mock dataset from the authors of \citet{Sohn_2922} to validate our pipeline and optimise our hyperparameters. The mock data are generated by adding features to the mean of BOBYQA best-fit power law spectrum to CamSpec PR3 unbinned TT, TE and EE data in (30,2500), (30, 2000) and (30, 2000) $\ell$ range respectively in addition to lowT (commander) and lowE (simall). These features were generated using five parameters which are not explicitly included in our grammar. The 100 realisations were generated assuming cosmic variance limited $\chi^2$ distribution for $C_{\ell}^{TT}$ in $\ell < 30$ range. For all other data, we draw $C_\ell$s from a multivariate normal distribution with the CamSpec covariance matrix.  

These tests serve two purposes:
1. Compare the results of the GA to the known ``true'' values of the mock dataset to see how well the algorithm found the optimal solution. 
2. Test the robustness of the GA by introducing variations or noise into the mock dataset and seeing how well the algorithm is able to handle these variations. 

\begin{figure*}
    \centering
    \includegraphics[width=0.95\textwidth]{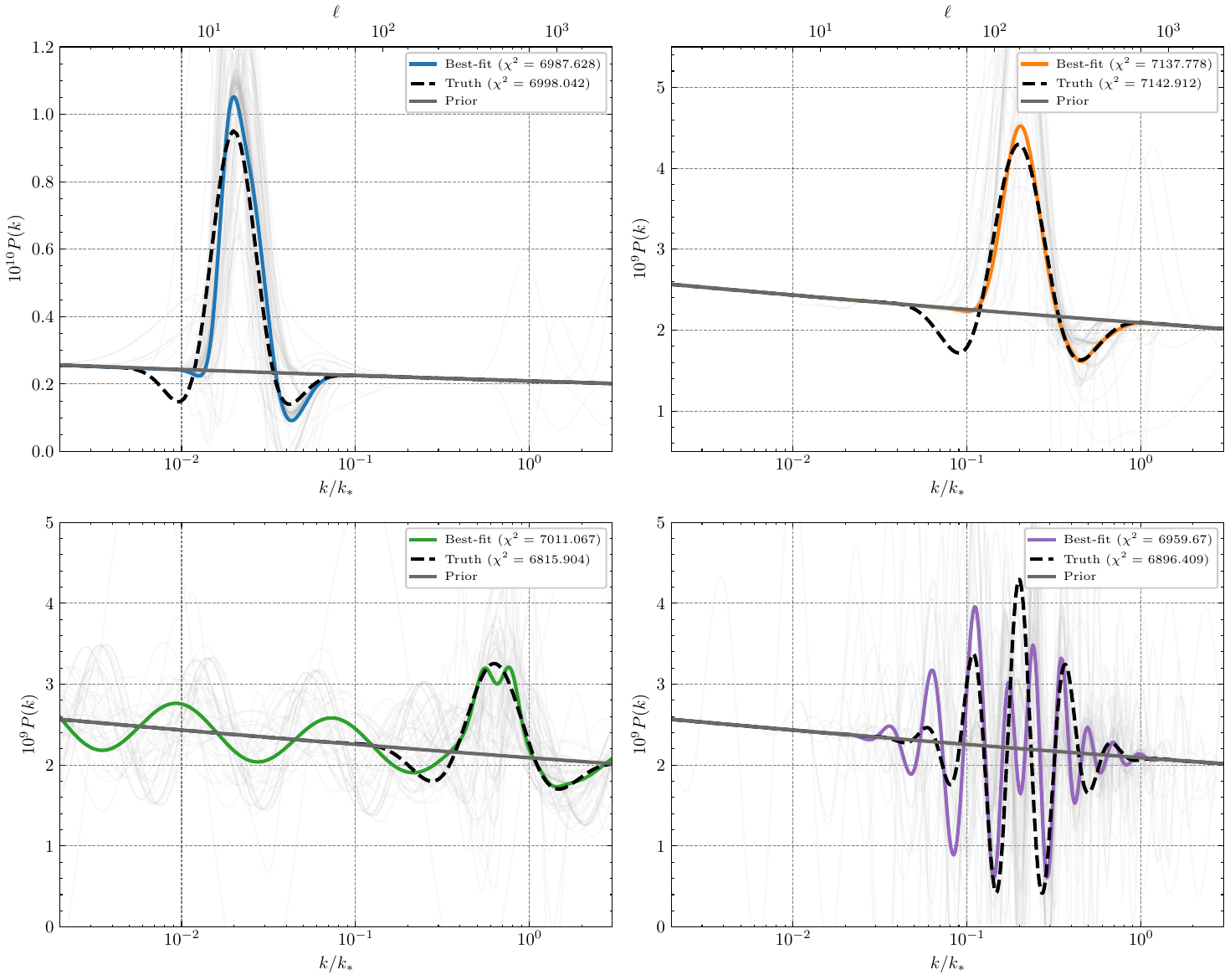}
    \caption{Samples from the final generation of GA explorations around power law with $N_{\mathrm{pop}}=100$ evolved till $N_\mathrm{gen}=500$ for four different feature mocks. The thick grey line shows the power law prior; the dashed black line shows the true feature used to create mock data, and the solid colour line shows the best-fit qualitatively detects the true feature in the mock. Additionally, silver lines depict the population of the final generation.}
    \label{fig:runVI_mockcsamples}
\end{figure*}

Due to the computational complexity inherent in our pipeline, we were constrained to conduct tests on a limited scale; we assessed our methodology's performance using five realisations of each of the four mock features.
In these tests, a single chain with an initial population of 100 individuals was run for 500 generations. The outcome of one of these runs is illustrated in Fig.~\ref{fig:runVI_mockcsamples}. It is important to note that the mock features employed in these tests were generated using five-parameter feature models, which were not explicitly included in our grammar.
Nevertheless, our pipeline exhibited a broad capacity to detect these features, albeit with slight variations in amplitude and the range of wave numbers corresponding to the true underlying features. Notably, in three instances, the GA identified the presence of a solitary feature in the mock and utilised the null function from our grammar to replace the other terms in the feature generation equation, given by Eq.~\eqref{eq:feature_gen}. 

\begin{figure}
    \centering
    \includegraphics[width=0.48\textwidth]{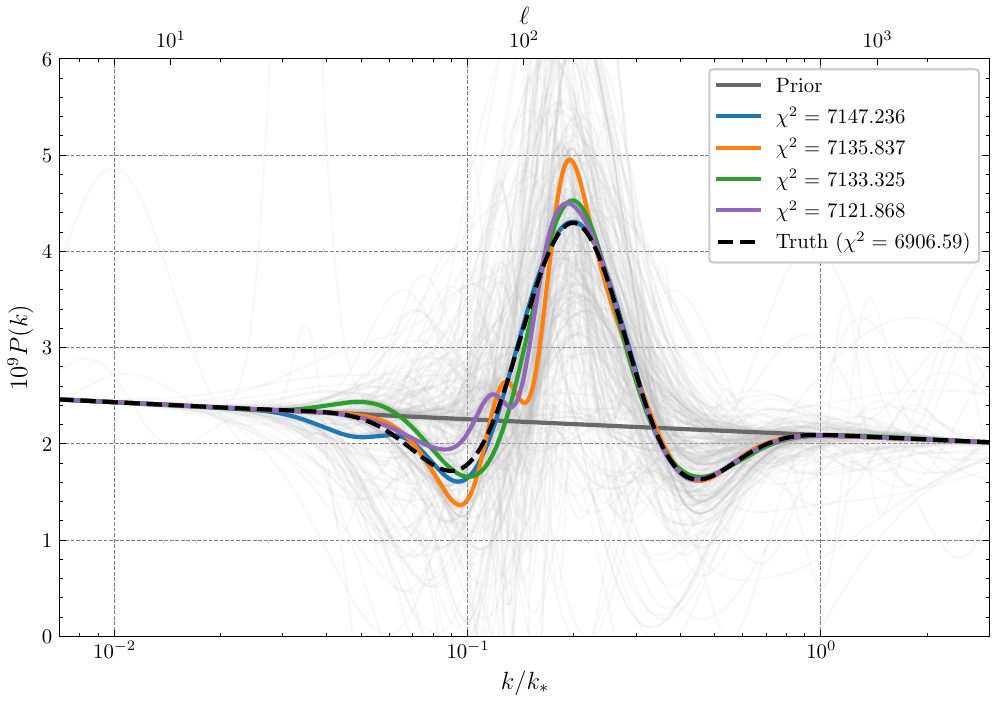}
    \caption{Best-fit power spectra for a featured mock with $N_{\mathrm{pop}} = 100$ and $N_\mathrm{gen} = 500$. Individual coloured lines correspond to four independent runs with different random seeds for the initial conditions. The grey line shows the power law prior; the dashed black line shows the true feature used to create mock data. We find best-fit power spectra from different chains and qualitatively find true features irrespective of the initial population. }
    \label{fig:runVI_mock1_4chains}
\end{figure}

Furthermore, we conducted an additional set of tests by running four supplementary chains with different seeds to investigate the impact of the initial population while fixing the mock dataset. GA consistently reproduced the shape and position of the features across all cases, demonstrating the robustness of our approach, as seen in Fig.~\ref{fig:runVI_mock1_4chains}.

Overfitting is a common pitfall where the model becomes too complex and excessively fits the noise in the data rather than the underlying signal. To mitigate this issue in the case of GA, we take several steps to mitigate the risk of overfitting. Firstly, we limit both the length and maximum number of generations of the GA to prevent extra tailoring of features. Additionally, we employ Bayesian Model selection after post-processing (see Appendix) to assess whether the data warrant additional degrees of freedom proposed by GA. Secondly, one can use  $\chi^2$ statistics using noise simulations of the data to quantify overfitting in terms of the probability to exceed (PTE) \citep{PlanckCollaboration2018power} but this is computationally expensive with GA and out of scope for this study.

Our results using simulations indicate that we can recover various forms of local features in the primordial spectrum. However, to assign significance to such recovered features (addressing the issue of noise-fitting) and, in particular, dealing with the real data, we need to perform some comprehensive analysis that is beyond the scope of this work. We should clarify here that in this work, we are not performing any model selection or model comparison, and we employ our approach to provide valuable insights into credible and plausible forms of the primordial spectrum that are consistent with CMB observations.

\section{Results}
\subsection{Data}

In this work, we utilise the latest CMB power spectra measurements from the Planck satellite, specifically the temperature and polarisation data sets provided by the Planck Collaboration \citep{PlanckCollaboration2018inflation}. We make use of two likelihoods, namely Plik and CamSpec, to analyse the data.

(i) \textit{plik}:  We use the unbinned version of the \texttt{plikTTTEEE+lowl+lowE+lensing} data set. The high-$\ell$ likelihood, referred to as \texttt{Plik}, provides power spectra in the temperature range $30 \leq \ell \leq 2509$ and polarisation range $30 \leq \ell \leq 1997$ by utilising cross-spectra between 100-, 143-, and 217-GHz maps \cite{PlanckCollaboration2018inflation}. On the other hand, the low-$\ell$ likelihoods (\texttt{Commander} and \texttt{SimAll}) cover the multipole range $2 \leq \ell \leq 30$ for the TT and EE spectra. Additionally, the CMB lensing likelihood provides measurements of the lensing trispectrum, estimating the power of the lensing potential $C^{\phi \phi}_{\ell}$ over $8 \leq L \leq 40$. These likelihoods are accessed via the $clik$ library, which provides the log-likelihood values used in our analysis. We refer to this data combination as ``Planck'' throughout the paper. The background cosmological parameters, nuisance parameters, and the tensor-to-scalar ratio $r$ are fixed to specific values obtained from the BOBYQA best-fit of \texttt{plikTTTEEE+ lowl + lowE + lensing}, with $r$ set to 0.

(ii) \textit{CamSpec}: In CamSpec analysis, we utilise the unbinned and co-added $C_{\ell}$s obtained from the CamSpec 12.6 version of the likelihood. CamSpec employs a correction procedure for each TE and EE cross-frequency spectrum, applying a fixed dust and temperature-to-polarisation leakage template before co-adding the spectra to form the EE and TE components of its data vector, see \cite{Efstathiou2021AMaps}. The CamSpec likelihood incorporates Planck's 100-, 143-, and 217-GHz maps, which were processed using the NPIPE pipeline. To further constrain the features in the low-frequency region ($\ell \in $ [2,29]), we include \texttt{lowTT}, \texttt{lowTE} and \texttt{lowEE} data from the Planck Legacy Archive (PLA) with a diagonal covariance matrix defined by uncertainties due to cosmic variance. The background cosmological and nuisance parameters are fixed to the values obtained from the BOBYQA best-fit of CamSpecTTTEEE+ lowl + lowE with $r = 0 $, ensuring consistency across our analysis. 

\subsection{Planck Runs}
 
\begin{figure*}
        \subfloat[Evolution of the $\chi^2$ as a function of the generations.]{%
            \includegraphics[width=.43\linewidth]{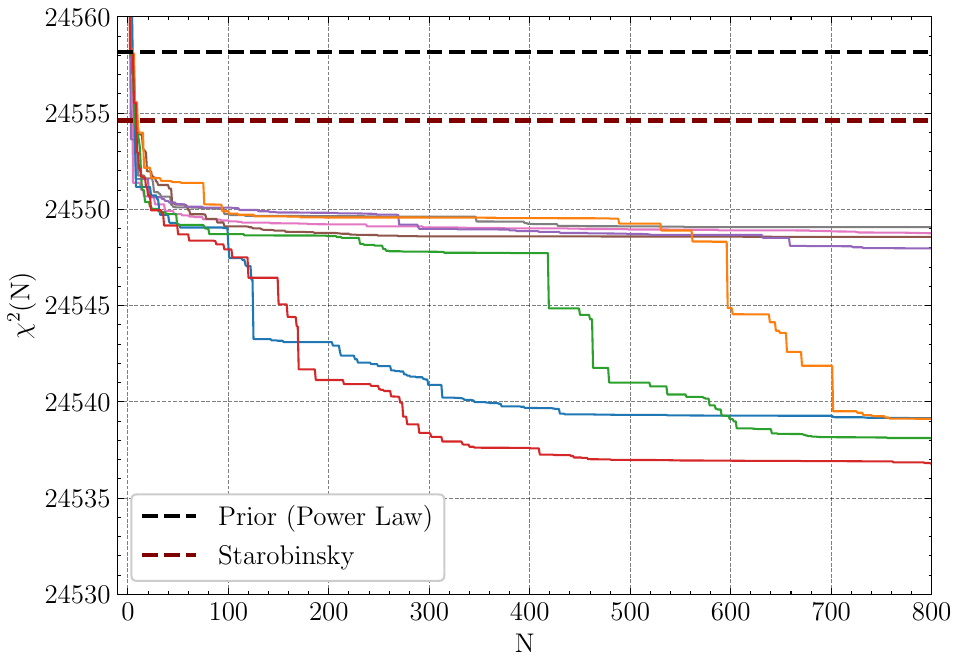}%
            \label{subfig:run6a}%
        }\hspace{0.1cm}
        \subfloat[Samples from the final generation.]{%
            \centering
            \includegraphics[width=.43\linewidth]{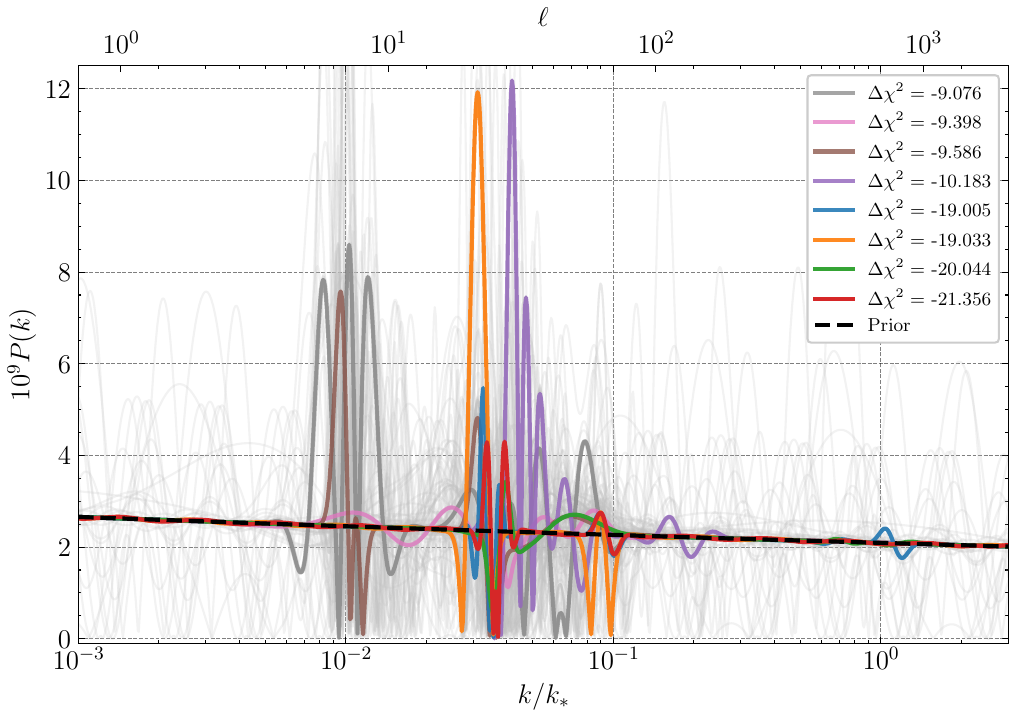}%
            \label{subfig:run6b}%
        }\hfill
        \subfloat[Residuals from the best-fit Planck Data.]{
        \centering
        \includegraphics[width=.9\linewidth]{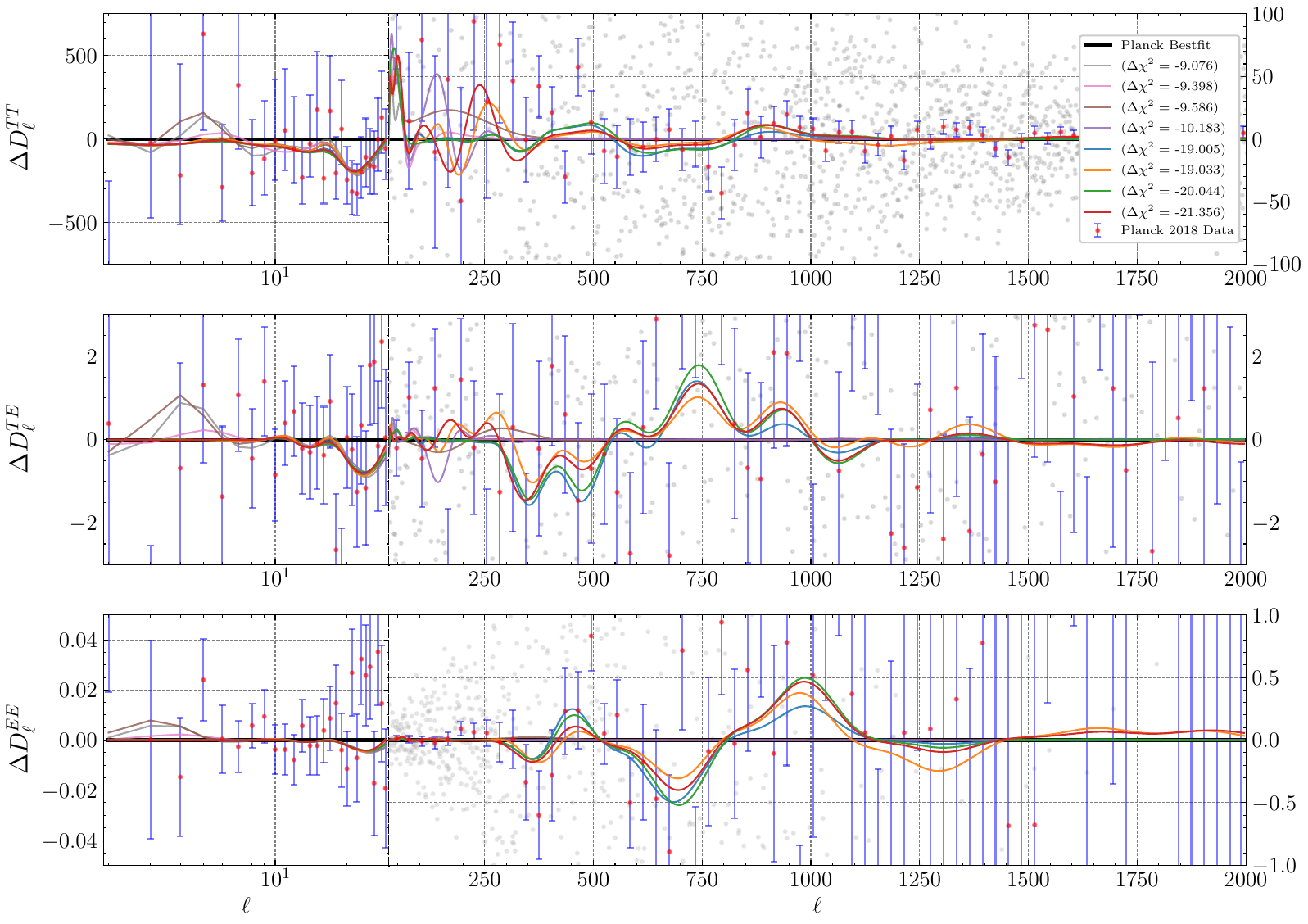}
        \label{subfig:res_planck}
    }
    \caption{Results from the GA search around a power-law power spectrum with Planck data after 800 generations with a population of 100. The individual coloured lines correspond to eight independent runs with different random seeds for the initial conditions. In the top left, we show the $\chi^2$ evolution of the best-fit individual in the population as a function of the number of generations. The best-fit $\chi^2$ from the power law and Starobinsky model are shown with dotted black and brown lines, respectively. We plot the best-fit power spectra from different chains in the top-right. The grey lines depict the rest of the population at the final generation, visually representing the distribution of individuals at the end of the evolutionary process. Finally in bottom subplot, we illustrate the residuals in $D^{TT}_{\ell}$, $D^{TE}_{\ell}$ and $D^{EE}_{\ell}$. The binned and unbinned data are shown in red and grey , respectively.  }
        \label{fig:plack_2018_results}
        
\end{figure*}

We apply our method to Planck 2018 data using the \textit{plik} likelihood as specified in the previous section.  Fig.~\ref{fig:plack_2018_results} shows our results for eight chains with an initial population of 100 after 800 generations. In the top-left Fig \ref{subfig:run6a}, we show the fitness of the best individual (combined $\chi^2$ from high-TTTEEE lowTT, lowEE 
and lensing likelihoods) as a function of generations. The horizontal black dotted and brown dotted lines are the best-fit of the power law model, and the Starobinsky model to Planck Data is provided as a reference. The Fig. \ref{subfig:run6b} shows the best-fit features in $P(k)$ space, where the best-fit individual from each chain is plotted in solid bold lines, and other individuals in the final generation are plotted in grey lines. The best-fit power law is shown as a black dotted line. 

To further investigate the spectra obtained from the GA, we plot the residual $D_{\ell}$ with respect to the best-fit power law in Fig \ref{subfig:res_planck}; points with error bars are binned Planck data, whereas grey points represent unbinned Planck data. In the TT spectra, we see features produce considerable shifts in range (20 < $\ell$ < 500), whereas in TE and EE spectra, we see the GA fitting peaks and troughs in the range (300 < $\ell$ < 1200). In addition to the residuals, in Table \ref{tab:delta_chi2_plik}, we show the breakdown improvement in $\chi^2$ between different likelihoods and further breakdown high-TTTEEE into high-TT, high-TE and high-EE assuming non-diagonal components do not contribute significantly. All chains improve greatly upon lowTT and high-TTTEEE data but at a small cost of low-EE data. While all chains address the low-$\ell$ anomaly, we also note that some chains perform much better than other ones with respect to high-$\ell$ data. This motivates us to apply the GA pipeline to high-$\ell$ data only, a procedure which we present in greater detail in App.~\ref{sec:app}.

The best-fit primordial power spectrum given analytically by the GA with the full Planck data is expressed in Eq.~(\ref{eqn:plik_bf}) below and features a combination of a global oscillation and local ones:

\begin{multline} 
F(x) =  0.014 \sin{\left(9.053x - 14.542 \right)} \\
-0.027 \sin{ \left( 8.777x-7.741 \right)} e^{-\left(8.777x-7.741 \right)^2}   \\
-0.565 \left(9.564x+22.498\right) \cdot  \cos{\left(9.564x+22.498\right)}  e^{-\left( 9.564x+22.498 \right)^2} \\
- 1.459  \cos{\left( 4 \cdot \left(9.296x+30.804 \right)\right)} e^{-\left(9.296x+30.804\right)^2},  \label{eqn:plik_bf}
\end{multline}
where $x = \log\left(\frac{k}{k_*}\right)$ as previously defined. The frequency of the global feature, in $\log k$ space, is $\omega_\mathrm{log} \sim 9$, which is of the same order of magnitude as the one found in specific searches for resonant features with the conventional Monte Carlo approach (see, e.g., Ref.~\cite{Braglia:2022ftm}), as well as in a previous application of GA to inflation, but at the level of the scalar potential~\cite{Kamerkar_2022}.

\begin{table*}
\caption{Summary of the best $\Delta \chi^2$ ( = $\chi^2_{\mathrm{GA}}$ - $\chi^2_{\mathrm{prior}}$) values for different Planck 2018 likelihoods across eight chains. The best chain has a $\Delta \chi^2$ value of -21.356, indicating a significant improvement in the fit \texttt{high-TTTEEE} and \texttt{lowl} although at a small cost of \texttt{lowE}}.
\centering
\begin{tabular}{|c|c|c|c|c|c|c|c|c|}
\hline
{Chain Index} &  low-TT &  low-EE &  high-TTTEEE &  lensing &  high-TT &  high-TE &  high-EE &   Total \\
\hline
0 &  -8.366 &   0.799 &     -1.757 &   -0.262 & -0.968 &  0.294 & -0.687 &  -9.586 \\
1 &  -8.233 &   0.793 &     -1.878 &   -0.080 & -0.002 & -0.453 & -1.835 &  -9.398 \\
2 &  -8.418 &   0.998 &     -1.616 &   -0.040 & -0.175 & -0.595 & -1.252 &  -9.076 \\
3 &  -7.580 &   0.528 &    -14.111 &   -0.193 & -5.092 & -5.928 & -6.703 & -21.356 \\
4 &  -7.796 &   0.608 &    -11.660 &   -0.157 & -3.095 & -7.435 & -6.746 & -19.005 \\
5 &  -7.336 &   0.509 &    -13.027 &   -0.189 & -3.658 & -8.342 & -6.527 & -20.044 \\
6 &  -7.261 &   0.641 &     -3.308 &   -0.254 & -0.757 & -1.314 & -1.232 & -10.183 \\
7 &  -8.197 &   0.821 &    -11.545 &   -0.113 & -2.800 & -6.600 & -5.294 & -19.033 \\
\hline
\end{tabular}
\label{tab:delta_chi2_plik}
\end{table*}

\subsection{CamSpec Runs}

\begin{figure*}
        \subfloat[Evolution of GA]{%
        \centering
            \includegraphics[width=.48\linewidth]{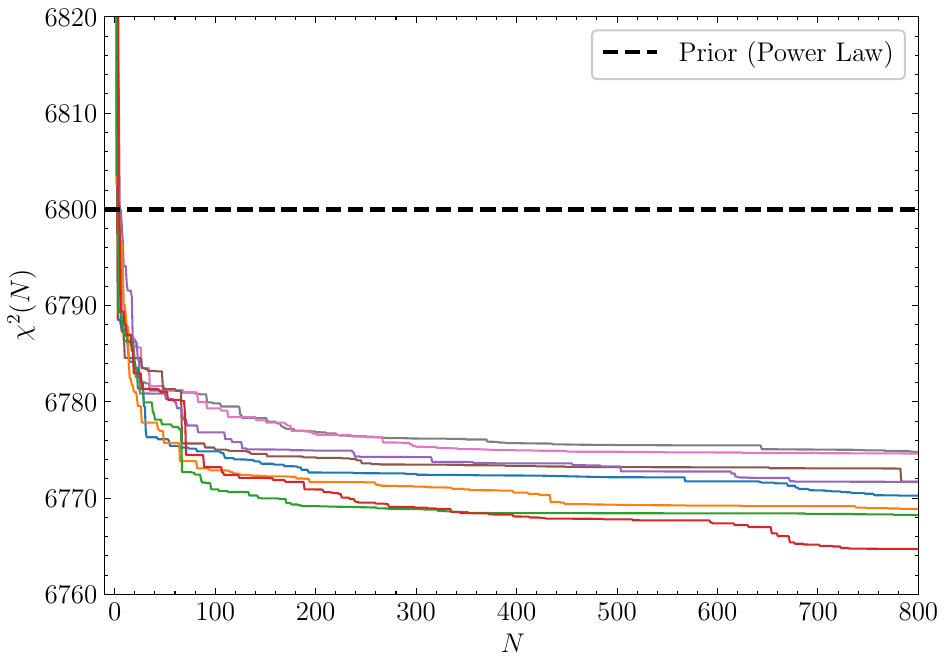}
            \label{subfig:run7a}%
        }\hfill
        \subfloat[Samples from Final Generation]{
        \centering
            \includegraphics[width=.48\linewidth]{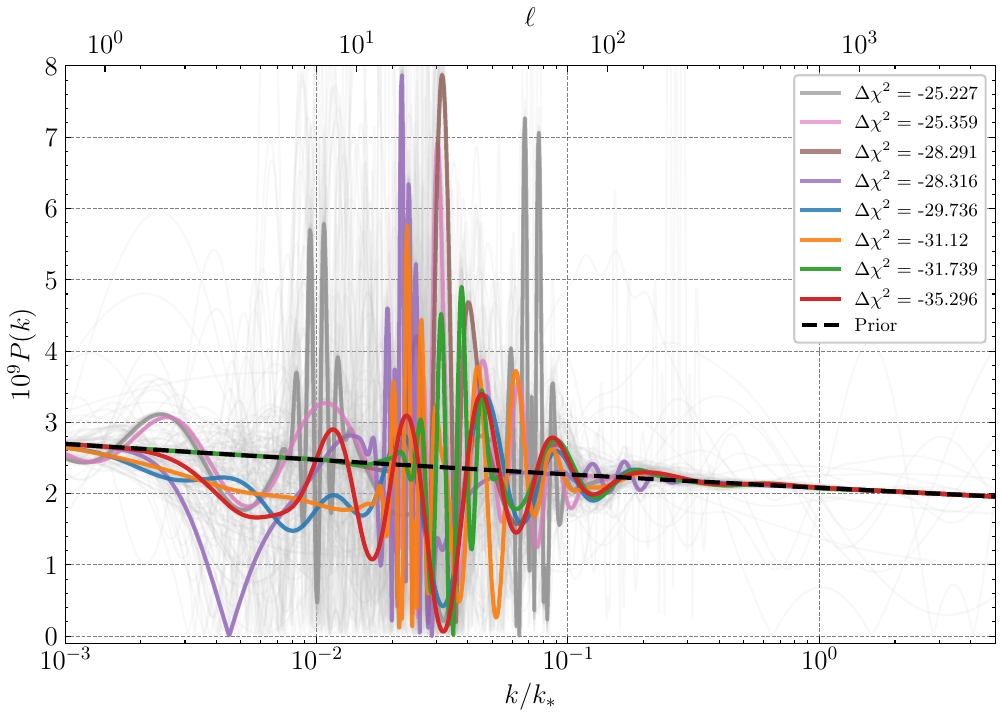}%
            \label{subfig:run7b}%
        }\hfill
        \subfloat[Samples from Final Generation]{
            \centering
    \includegraphics[width=.9\linewidth]{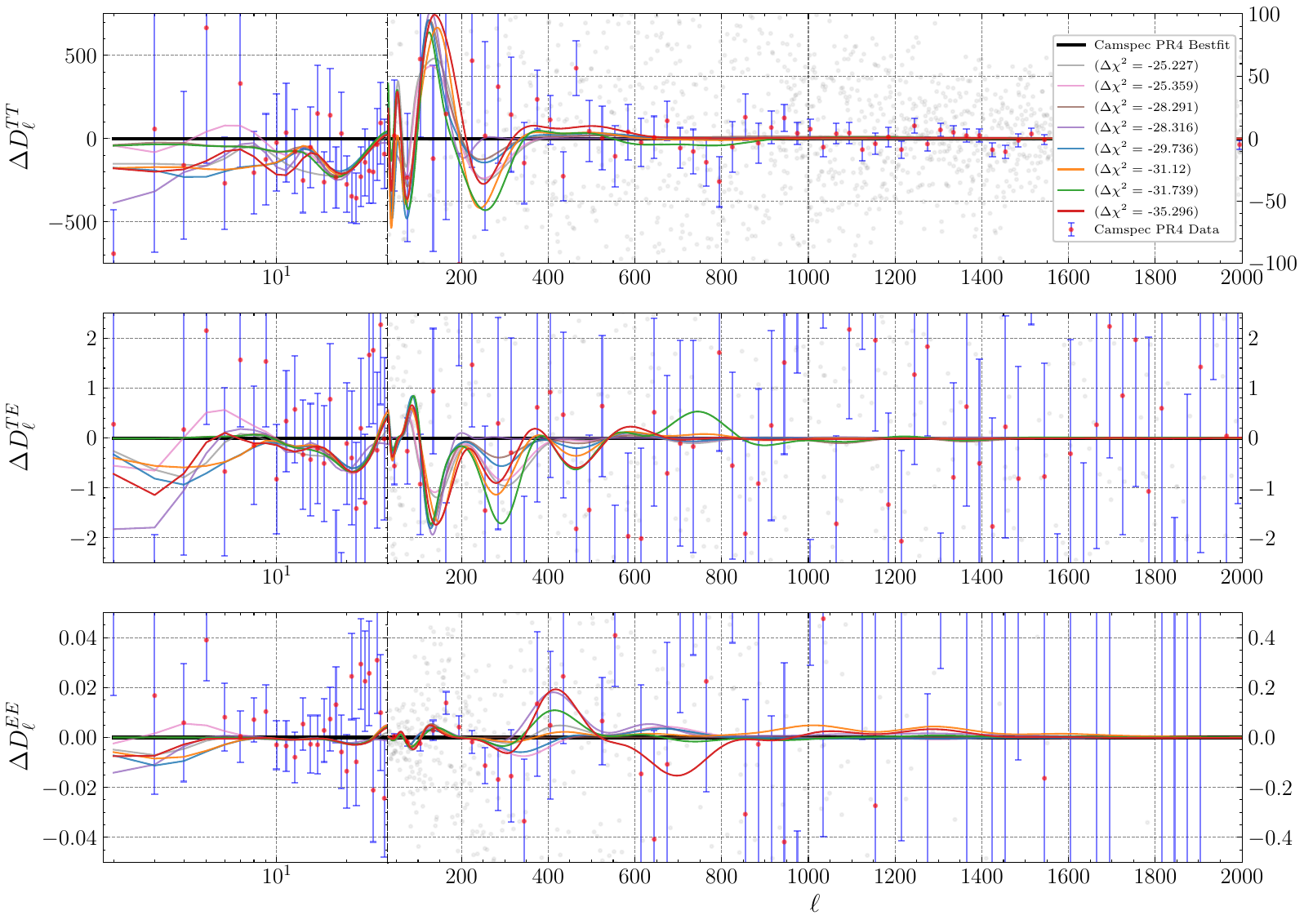}
    
    \label{subfig:res_camspec}
        }
        \caption{Results from the GA search using CamSpec data after 800 generations with a population of 100. Each solid-coloured line corresponds to one of the eight independent runs with different random seeds for the initial conditions. The top left plot displays the evolution of the $\chi^2$ value for the best-fit individual in the population as the number of generations increases. The dotted black line represents the best-fit $\chi^2$ from the power law model. In the top-right plot, we showcase the best-fit power spectra obtained from different chains and grey lines illustrating the population distribution in the final generation. In the bottom plot, we illustrate the residuals in $D^{TT}_\ell$, $D^{TE}_{\ell}$, and $D^{EE}_{\ell}$ are illustrated.}
        \label{fig:camspec_pr4_results}
\end{figure*}

Similar to the previous section, we perform our analysis on CamSpec PR4 data by taking co-added $C^{TE}_{\ell}$s and $C^{EE}_{\ell}$s along with $C^{TT}_{\ell}$ obtained following the method of \citet{Planck_2016b} for the maximum likelihood solution. The nuisance and background parameters, in addition to $A_s$ and $n_s$, were fixed to the values taken from the CamSpecTTTEEE+lowl+simallE best-fit parameters. In Fig.~\ref{subfig:run7a}, we show the evolution of total $\chi^2$ for eight chains through generations. We observe all eight chains outperform the power law shown in the black dotted line. Next, Fig.~\ref{subfig:run7b} shows best-fit individuals from each chain in the $P(k)$ space. The GA finds considerable deviations from the best-fit power-law spectrum in the $\ell < 1000$ region.

In Fig.~\ref{subfig:res_camspec}, we plot the $D_{\ell}$ residuals for TT, TE and EE spectra, which makes it apparent where a significant share of improvement in $\chi^2$ is coming from lowT and EE data. We also see our pipeline trying to fit the peak in EE residual space around $\ell \approx 400$. In Table \ref{tab:delta_chi2_camspec}, we quantify the improvement in $\chi^2$ from the best-fit individuals. All eight chains have more than 25 $\chi^2$ improvements, with the best one $\Delta \chi^2 \approx -35$ compared to the fiducial power law model. %One important fact to note is that two of these features improve all spectra except lowEE. 
The analytically best-fit spectrum obtained from the GA with the CamSpec run is given by: 
\begin{multline}
    F(x) = 0.019 \cdot \left(0.648 x + 1.657 \right) \cdot \sin{\left( 0.648 x + 1.657 \right)} e^{-\left(0.648 x + 1.657  \right)^2} \\
    + 0.065 \cdot \cos{\left( 8 \cdot \left(0.694 x + 1.800 \right)\right)} e^{ - \left(0.694 x + 1.800 \right)^2} \\
    - 0.703 \cdot \cos{\left( 8 \cdot \left(1.127 x + 3.877 \right)\right)} e^{ - \left(1.127 x + 3.877 \right)^2} \\
    - 1.143 \cdot \left(1.292 x + 5.723 \right) \cdot \sin{\left(1.292 x + 5.723 \right)}  e^{- \left(1.292 x + 5.723 \right)^2}. \\
    \label{eqn:bf_camspec}
\end{multline} 

Our analysis reveals that in both the CamSpec and Planck runs, a significant portion of the improvement is attributed to fitting the lower power in the spectra around $\ell \approx 20$ along with the small periodic deviations in the EE spectra at higher multipoles in CamSpec runs, which further improve the fit to the data as compared to the Planck runs. However, we do not observe the features obtained by the Planck run in the TT spectra within the multipole range of $\ell = 750-1000$ in CamSpec runs. The global feature found with the Planck run is also consistent with the first application of GA to inflation~\cite{Kamerkar_2022}, although the best-fits found in this previous work did not address the low-$\ell$ anomaly at all.Furthermore, the oscillations in between $k/k_*=10^{-2}$ and $10^{-1}$ of the best chain (red line in Figure \ref{subfig:res_camspec}) are of similar frequencies and phase seen in the regularised MRL results \citep{Sohn_2922} on the CamSpec data.This concurrence highlights the robustness and reliability of our results and strengthens the validity of the observed features in the primordial power spectrum.

\section{Summary and Discussions}

Our study introduces a novel GA methodology to effectively explore and identify features within the primordial power spectrum. By incorporating a high degree of freedom that can be interpreted analytically, our approach enables efficient and targeted searches for these features. Investigating features is essential as they can provide valuable insights into the underlying physics of the early Universe, potentially alleviate cosmological tensions, and help us validate or rule out inflationary models. 

Through tests on mock data with known true features, we evaluated the performance of the GA pipeline. Despite not explicitly incorporating the specific five-parameter featured models used to generate the mock data in our grammar, the GA demonstrated its ability to broadly detect these features, although with some amplitude and $k$-range shifts. Further refinement is needed in the pipeline, particularly in improving the efficiency of computing CMB angular power spectra (\Cell) and its likelihood, to enable more accurate and comprehensive feature searches by incorporating variations in background and nuisance parameters.

Our study showcases the effectiveness of genetic algorithms in directly identifying features in the primordial power spectra using Cosmic Microwave Background observations.
By applying our pipeline to Planck and CamSpec PR4 data, we identify potential features using a simple motivated grammar made of both global and local features.
These features may be due, e.g., to transient deviations from the slow-roll dynamics during inflation or persistent small features atop a smooth primordial scalar potential. 
Another advantage of GA lies in their ability to provide best-fit analytical solutions that consist of combinations of individual features.
This allows us to evaluate the effects of these features individually, helping us to identify the source of improvement in fitness to data more precisely than other non-analytic methods.

By disentangling the effects of these individual features, we can construct a model with fewer parameters.
For instance, in one of our high-TTTEEE runs, which are further explained in the appendix,  we discovered a feature that significantly improved fitness to data, reducing the $\chi^2$ value by $\Delta \chi^2 \approx -12$. This finding can help us design a featured model with just three parameters (A, C, D), whose significance can be assessed by employing an MCMC analysis and calculating the Bayesian evidence.
This quantification will provide additional insights into the reliability and robustness of the features found by GA. We should emphasise here that in this work, we are not performing any model selection analysis, and we use GA mainly to suggest some plausible and viable form of the primordial spectrum that can fit CMB observations well.
In our future work, we plan to extend our research by incorporating more complex grammatical structures and evaluating their effectiveness on various CMB surveys, including SPT \citep{SPT3G:2014dbx}, ACT \citep{Aiola_2020}, and their combinations with Planck data. Moreover, we expect the straightforward adaptability of our methodology to forthcoming survey data, such as the Simons Observatory (SO) \citep{Ade_2019} and CMB-S4 \citep{CMBS4:2016ple}.

The regime of the applicability of the techniques developed in this work goes well beyond testing primordial features with current and future CMB probes.
Upcoming galaxy surveys such as DESI  \citep{levi2013desi}, EUCLID \citep{laureijs2011euclid}, and LSST \citep{LSST:2008ijt}, in addition to 21-cm tomographic observations, exhibit raw statistical power comparable to CMB measurements and would provide valuable complementary constraints \citep{Chen_2016_21cm,Chen_2016, Palma2017ConstraintsOI,Chandra_2022,chandra2022exploring}.
From a theoretical standpoint, it is crucial to emphasise the potential of our analysis methodology to explore deviations from scale-invariance, particularly in the tensor power spectrum. Our approach facilitates a model-independent investigation of such deviations through the primordial B-modes, with consideration for their potential detection in the forthcoming generation of CMB experiments, such as SO, CMB-S4, and LiteBIRD. %\textcolor{green}{
One could also naturally expand the scope of our analysis to include features atop a GW signal at different scales, such as the one recently detected by pulsar timing array experiments \citep{NANOGrav:2023gor,Antoniadis:2023rey, Reardon:2023gzh,Xu:2023wog}. Our approach to features would be equally effective also in testing primordial physics with future gravitational wave detectors at small scales.

\begin{table*}
\caption{Summary of the best $\Delta \chi^2$ ( = $\chi^2_{\mathrm{GA}}$ - $\chi^2_{\mathrm{prior}}$) values for co-added CamSpec data across eight chains. All eight chains have a $\Delta \chi^2$ value $<$ -25, with best-fit improving $\chi^2$ by 35.296, signifying a notable improvement compared to the power law model.}
\begin{tabular}{|c|c|c|c|c|c|c|c|}
\hline
{Chain Index} &   lowTT &  lowTE &  lowEE &     TT &     TE &     EE &   Total \\
\hline
0 & -15.065 & -2.526 &  2.468 &  2.074 & -4.919 & -11.768 & -29.736 \\
1 & -14.156 & -1.883 &  1.850 &  2.986 & -3.475 & -10.549 & -25.227 \\
2 & -17.218 & -3.917 &  1.426 & -0.460 & -2.256 &  -5.891 & -28.316 \\
3 & -15.416 & -2.848 &  1.769 &  3.425 & -4.389 & -17.837 & -35.296 \\
4 & -11.342 & -3.224 &  0.978 &  2.825 & -4.251 & -10.344 & -25.359 \\
5 & -12.445 & -2.750 &  1.175 &  2.544 & -5.105 & -11.709 & -28.291 \\
6 & -15.646 & -2.654 &  2.376 &  4.427 & -3.926 & -15.698 & -31.120 \\
7 & -12.114 & -2.534 &  1.367 &  2.547 & -2.990 & -18.015 & -31.739 \\
\hline
\end{tabular}
\label{tab:delta_chi2_camspec}
\end{table*}

\section*{Acknowledgments}
%\newline
The authors acknowledge the use of computational resources of the high-performance computing cluster Seondeok at the Korea Astronomy and Space Science Institute. We thank Dhiraj Hazra and Akhil Antony for the useful discussion and for providing BOBYQA runs of CamSpec data. We also thank Matteo Braglia for his comments and feedback. 

SN acknowledges support from the research project PID2021-123012NB-C43. MF and LP would like to acknowledge support from the “Ramón y Cajal” grant RYC2021-033786-I. The work of MF, LP, SN is partially supported by the Spanish Research Agency (Agencia Estatal de Investigación) through the Grant IFT Centro de Excelencia Severo Ochoa No CEX2020-001007-S, funded by MCIN/AEI/10.13039/501100011033. AS would like to acknowledge the support by National Research Foundation of Korea 2021M3F7A1082056, and the support of the Korea Institute for Advanced Study (KIAS) grant
funded by the government of Korea. A.S. would like to thank IFT for the hospitality during the early stages of this work. 

%%%%%%%%%%%%%%%%%%%%%%%%%%%%%%%%%%%%%%%%%%%%%%%%%%
\section*{Data Availability}
The data used in this work are available in Planck Legacy Archive (PLA), at \url{http://pla.esac.esa.int/pla/#cosmology}. Mocks are generated by the authors, as described in the text. Code and Mocks will be shared upon a reasonable request to the corresponding author.
 
% \newpage


\bibliographystyle{mnras}
\bibliography{ref} % if your bibtex file is called example.bib
%\nocite{*}

%\begin{thebibliography}{99}
%\end{thebibliography}

%%%%%%%%%%%%%%%%%%%%%%%%%%%%%%%%%%%%%%%%%%%%%%%%%%

%%%%%%%%%%%%%%%%% APPENDICES %%%%%%%%%%%%%%%%%%%%%

\appendix

\section{Runs with Planck high-TTTEEE data only }
\label{sec:app}
In this appendix section, we present the results of our genetic algorithm (GA) run using \texttt{Planck high-TTTEEE} data only.
The motivation behind this run stems from the observation that, although all Planck 2018 chains showed an improved fit to the dip at low $\ell$, only some of them also addressed the peaks in the high-$\ell$ angular power spectra. We aim to investigate whether the GA can systematically identify features that significantly enhance the fit to the specifically high-$\ell$ data, which possesses higher constraining power compared to the cosmic variance dominated low-$\ell$ region.

\begin{figure}
    \centering
    \includegraphics[width=\linewidth]{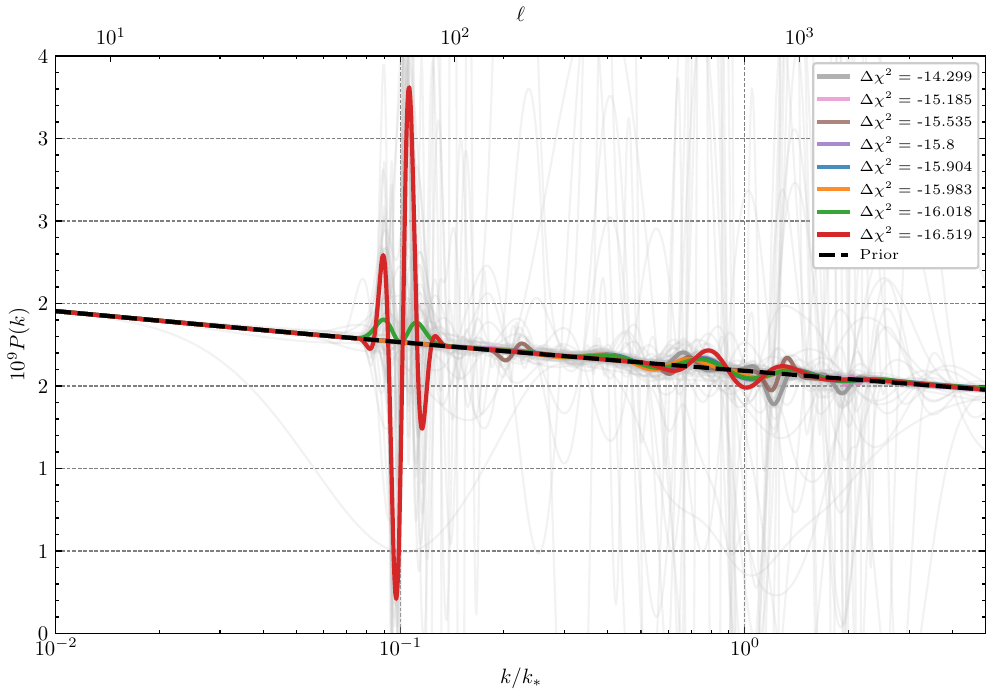}
    \caption{The outcomes of the GA exploration using \texttt{planck high-TTTEEE} data are showcased following 800 generations, with a population size of 100.}
    \label{fig:hiTTTEE_samples}
\end{figure}
% \newpage

% \vspace{7cm}
% \newpage
\begin{figure}
    \centering
    \includegraphics[width=\linewidth]{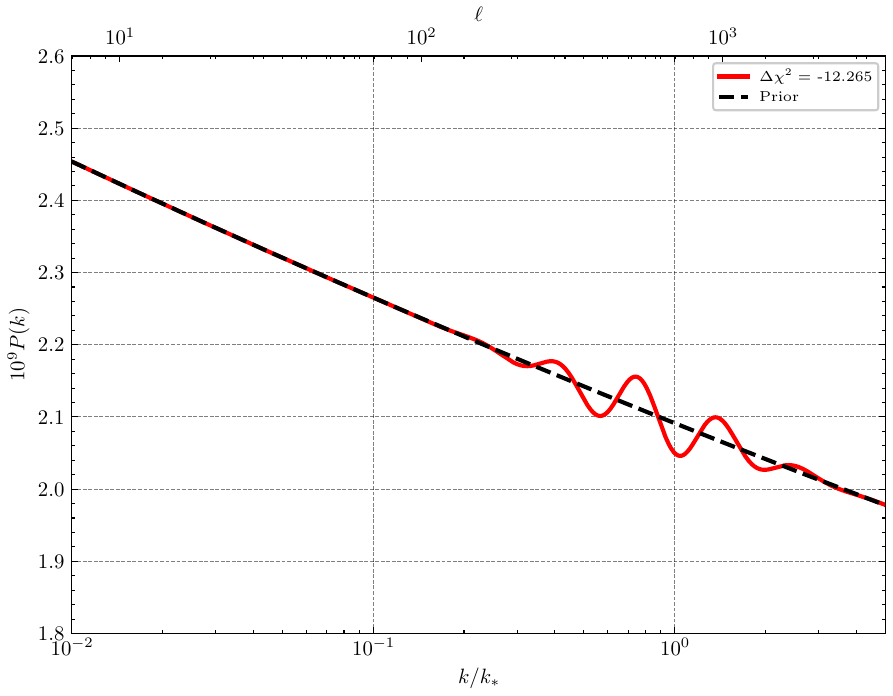}
    \caption{Best single feature obtained by scrutinizing the individual features in the best-fit chain. This single feature is local and is analytically expressed as $f_3 (x) = -0.021 \sin(8 \cdot(1.237 x + 0.159))e^{-(1.237 x + 0.159)^2} $ where $x = \log\left(\frac{k}{k_*}\right)$. }
    \label{fig:bf_sigle_feat_pk}
\end{figure}

% \vspace{7cm}
% \vspace{10cm}
\newpage
\begin{figure}
    \centering
    \includegraphics[width=\linewidth]{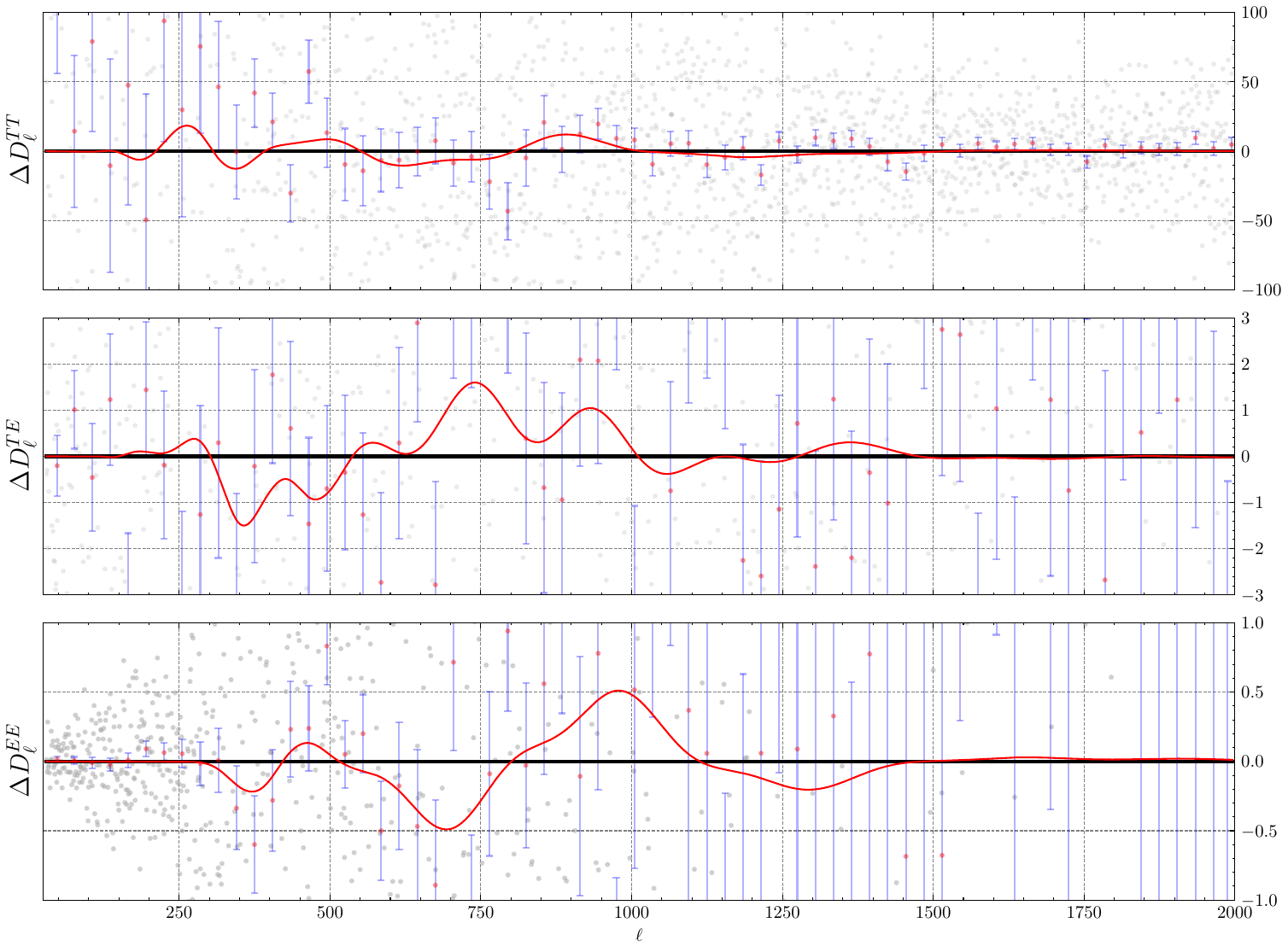}
    \caption{The residuals in $D^{TT}_\ell$, $D^{TE}_{\ell}$, and $D^{EE}_{\ell}$ for best single feature with respect to the power-law best-fit. }
    \label{fig:bf_sigle_feat_cls}
\end{figure}

% \columnbreak

% \newpage/

To accomplish this, we modified the prior position of the feature (D), narrowing it down from the original range of $ k/k_* \in [0.01, 5]$ to $ k/k_* \in [0.1, 5]$. The GA run was conducted with eight chains, maintaining all other parameter configurations identical to the Planck runs. The final generation from the various chains is depicted in Fig \ref{fig:hiTTTEE_samples}.
Remarkably, all eight chains exhibited a notable improvement in fitness to the data, resulting in an approximate reduction of $\chi^2$ by around -15. By strategically narrowing the prior on the position of the feature, we directed the GA's exploration to focus on the dips and peaks in the high-$\ell$ data only, resulting in substantial improvements in fitness to this feature.
As part of post-processing, we check all permutations on the features to identify which individual features were responsible for most of the improvement.

In this case, we found that one specific feature ($f_3$ in $f=\sum_{i=1}^4 f_i$), illustrated in Fig \ref{fig:bf_sigle_feat_pk}, contributed $\sim 75 \%$ of the total improvement of the fit to the data, as seen in Fig \ref{fig:bf_sigle_feat_cls}, with a $\Delta \chi^2= -12.265$.

\begin{table}
\caption{Evidence of Featured model with respect to the power law. Positive $\Delta \log Z$. }
\begin{center}

\begin{tabular}{|l|l|l|}
\hline
Model              & $\Delta \log Z$     & Strength             \\
\hline
Broad A prior   & -2.44  & Moderately Disfavoured \\
\hline
Narrow A prior  & 0.05 & Slightly Favoured \\
\hline
% \end{center}
\end{tabular}
\end{center}
\label{Tab:Bayesian_evidence}
\end{table}

\newpage

This motivated us to propose a three-parametric extension to the power law model, where we fix the shape but vary the amplitude (A), width (C), and position (D) of the feature. We tested this extension against the power law using the plik likelihood with a nested sampler. All standard parameters (except $\tau_{\mathrm{reion}}$) were allowed to vary within the prior range mentioned in \citep{PlanckCollaboration2018inflation}. We adopted uniform priors from Table \ref{Tab:prior}, with the exception of parameter A, which was allowed to vary in the range of [-1, 1], based on the reasonable assumption that the amplitude of the feature cannot exceed the amplitude of the power law. Additionally, we considered a case where parameter A is confined to the range [-0.1, 0.1]. The results from Table \ref{Tab:Bayesian_evidence} indicate that the featured model is not favoured over the power law but is consistent with current data and would require additional datasets to be ruled out.

This demonstrates the utility of GA and post-processing in identifying potential new models that can surpass the vanilla model with a minimal amount of additional degrees of freedom.

% Don't change these lines
\bsp	% typesetting comment
\label{lastpage}
\end{document}